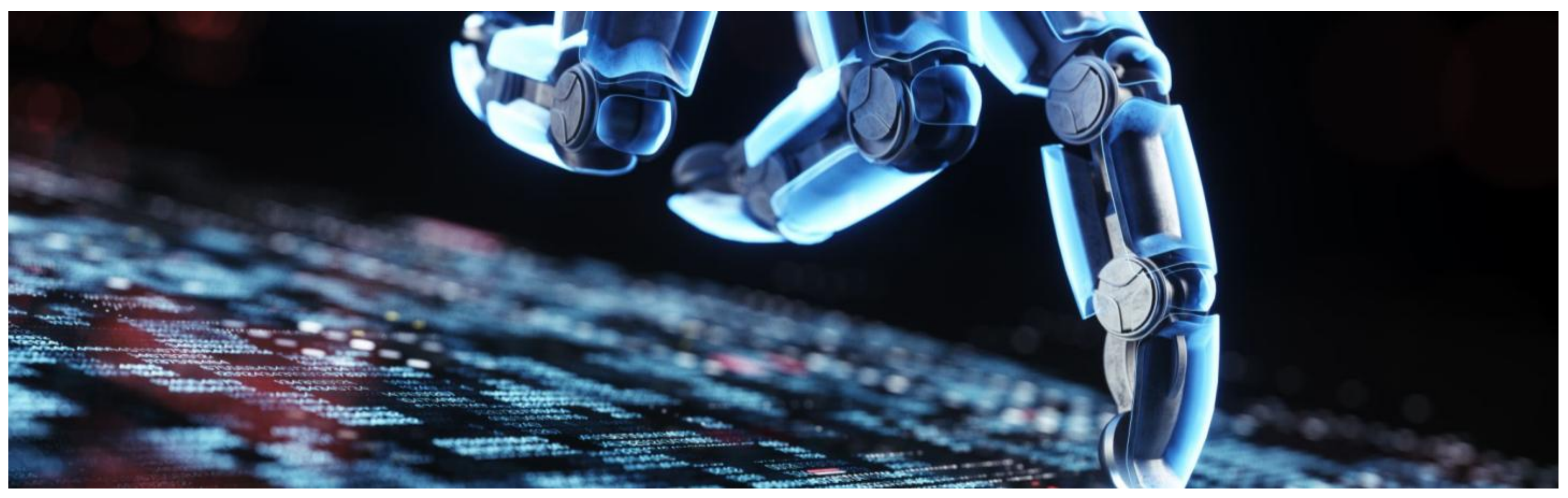

# Deepfake Technology Unveiled: The Commoditization of AI and Its Impact on Digital Trust

Abiraam Kesavarajah
100701559

Hewad Tahiri
100824648

Liam Cunningham
100821014

Rex Pallath
100847414

Tao Wu
1006899

Claudiu Popa
Project Leader
ORCID 1-9434-1807

***Abstract*—Driven by the commoditization of multimedia AI tools, deepfake technology has become a disruptive force in today's digital landscape. With the increasing accessibility of generative AI, tools for voice cloning, face-swapping, and synthetic media creation have advanced significantly, lowering both financial and technical barriers for their use. While these technologies present innovative opportunities, their rapid growth raises concerns about trust, privacy, and security. This white paper explores the implications of deepfake technology, analyzing its role in enabling fraud, misinformation, and the erosion of authenticity in multimedia. Using cost-effective, easy to use tools such as Runway, Rope, and ElevenLabs, we explore how realistic deepfakes can be created with limited resources, demonstrating the risks posed to individuals and organizations alike. By analyzing the technical and ethical challenges of deepfake mitigation and detection, we emphasize the urgent need for regulatory frameworks, public awareness, and collaborative efforts to maintain trust in digital media.**

***Keywords—artificial intelligence risks, content authenticity, deepfake fraud, deepfakes, digital trust, face-swapping, generative ai, media integrity.***

## I. INTRODUCTION

The rapid advancement of artificial intelligence (AI) has given rise to deepfakes, a powerful and controversial technology. Originally used for harmless purposes, deepfakes are increasingly being exploited for malicious activities such as misinformation campaigns, financial fraud, and social engineering attacks. However, the same technology also shows potential for positive applications, including enhancing creativity, improving accessibility, and advancing educational content. As deepfake technology becomes more accessible and commodified, it is crucial to explore the dual nature of its impact on society. This white paper focuses on the use of deepfake video, audio, and generative AI and examines the ethical, social, and technological implications of deepfake technology, weighing its potential for both good and bad.

## II. SOFTWARE & COST ANALYSIS

### A. *Deepfake Analysis*

In the last decade, the explosive rise of AI has led to an enormous boom of technological leaps in deep learning and an increase in both the complexity of these deep learning algorithms and their increased commercial use. It is precisely this new commercial availability, alongside its rapidly improving quality, that makes deepfakes such a prevalent threat in the modern world. As these tools become more advanced it becomes more likely perpetrators will use them for their malicious purposes [1]. Deepfake content can be put to numerous illegal or malicious uses, from spreading dangerous misinformation, and the production of non-consensual AI pornography, to use in large-scale theft or fraud.

The looming threat of deepfake fraud poses a significant threat to both individuals and large organizations, with fraudulent deepfake scams resulting in potentially millions of dollars in fiscal losses as well as significant reputational damage. One such case in 2019 involved a Californian woman defrauded of $300,000 US dollars by criminals posing as romantic partners on a dating site, using AI deepfakes to impersonate US Navy Vice Admiral Sean Buck [2]. The high-profile figures these criminals imitate often remain unaware of their likeness or name being used in these fraudulent actions. They may suffer the consequences of their doppelganger's actions, either legally or reputationally. Cases such as the one previously mentioned can have devastating effects on an individual or a select few individuals, but some deepfake scams can jeopardize the activities or functions of entire organizations. A particularly high-profile case involved perpetrators using deepfake voice-calling technologies to imitate a director of the United Emirates Bank, defrauding the bank of the equivalent of $35 million US dollars [3]. These are only a few of the social engineering schemes being utilized by modern criminals, and with every passing year, they become more elaborate and more convincing.

However, there is a monetary cost associated with using powerful deepfake replication software, as well as a time cost to learn and manage these systems. In our efforts to mimic a fraudulent scam supported by deep learning AI, we have gathered the costs of performing simple scams and determined just how easy it truly is to create such a malicious scam.

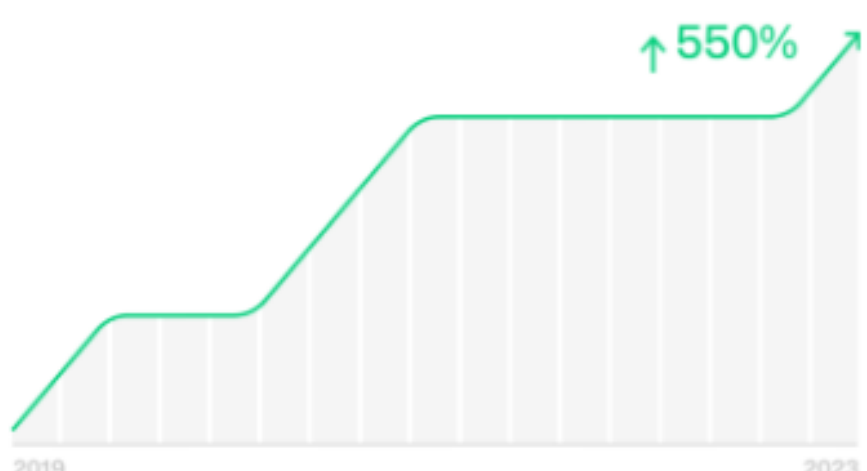

Fig. 2.1. The number of deepfake videos available online [4].

## B. *Software Analysis*

The inherent cost of running successful deepfake campaigns or scams depends on the scope and scale of the intended scam. To perform small-scale deepfake scams, such as impersonating a celebrity or individual to deceive a single person, a threat actor could do so with negligible monetary or technical investment. However, performing large-scale fraud, such as impersonating a CEO on a live video call, is incredibly technically challenging and would require significant monetary investment in technology and software. Additionally, threat actors can outsource the creation of deepfakes; for example, the site Fiverr hosts numerous individuals willing to create deepfake content for as little as $5. Some of these individuals are even willing to fabricate videos of specific people, such as exes, classmates, or coworkers [5]. For our purposes, we will discuss the software used in our mock scam and the costs associated with it.

## C. *Runway Summary*

Founded in 2018, the Runway AI Company has been a prevalent provider of commercial deepfake software for over half a decade. The company itself claims to be dedicated to the constant improvement of its AI tools and research into the AI field, going as far as to say that it is "laser-focused on applying research to product development and therefore broadening access to AI-backed experiences for our users" [6].

The Runway Unlimited plan provided us with numerous benefits and tools that offset its steep cost and provided a strong backbone to our production software. Notably, Unlimited Mode provides unlimited generation using Runway's powerful Gen-3 Alpha Turbo AI to generate our base videos, as well as 2,250 credits to spend on improved efficiency and speed for generating our videos [6]. The core functions of Runway are integral to our deepfake production, as Runway provides tools for both developing and editing our base video using their generative AI, as well as a custom tool for lip-syncing our generated faces to our script to make the video as seamless and smooth as possible, which works in tandem with our second piece of deepfake technology, Rope.

## D. *Rope Summary*

Rope is an open-source deepfake and face-swapping GUI built on top of Roop, which is itself an open-source face-swapping program created in 2023 [7]. Being an open-source product, Rope is constantly being updated and improved, with users and members of various online communities dedicated to suggesting and even testing updated or improved software [8].

Rope provided us with a few especially important functionalities for our deepfake imitation. First and foremost, Rope was our primary face-swapping software, switching the face of our victim onto the AI-generated videos made by Runway. While Runway was our primary tool for fixing lip-syncing, most of our cosmetic fine-tuning was also done by Rope. Rope can save and edit specific frames of its enhanced content, and with powerful editing tools such as the Yolov8 face detection model, it provided us with many small cosmetic corrections and enhancements to make our falsified video appear more realistic to the untrained onlooker [8]. In addition, Rope has incredibly efficient memory usage, allowing it to run easily and quickly even on older machines, making it a prime piece of software for threat actors to employ, as it can run on cheap hardware and be accessed for free.

*E. ElevenLabs Summary*

Founded in 2022, The ElevenLabs ~~company~~company claims to be dedicated to the research and enhancement of AI voice cloning and editing technologies and claims to produce authentic, versatile, and contextually aware AI voices [9]. While ElevenLabs has a free plan to utilize its powerful software, we selected the Creator plan for higher-quality audio generation to recreate the most realistic version of our victim's voices.

ElevenLabs provides its users with some key tools that make realistic cloning and generation of AI voices easy for even those who lack the typical knowledge of voice cloning software. To get the most realistic and consistent voice required to conduct our deepfake video, we were able to draw on two specific functions of the ElevenLabs software. The foremost use of ElevenLabs was its voice cloning feature, allowing us to feed it audio clips of the victim's voice to produce an accurate cloned model of the victim's speech patterns and overall voice. Secondly, the ElevenLabs voice changer feature allows the editing of our voices to sound like the victims, enabling us to mimic specific phrases the AI may find difficult or reproduce the victim's voice in a theoretical live scenario.

*F. Cost Analysis*

TABLE I. SOFWARE COSTS (CAD)

| Software Name | Final Cost |
|---|---|
| Runway | $135.87 |
| Rope | $0 |
| ElevenLabs | $17.74 |

As previously discussed, the price of creating deepfake content has rapidly decreased in the last half-decade alongside the sharp decline in the barrier of technological skill and knowledge required to create them. In our efforts to mimic a deepfake scam, we spent a total of $153.61 CAD to subscribe to both Runway and ElevenLabs for AI video generation, face swapping, and voice cloning technologies. However, stepping back into 2019, only a handful of years ago, it was far more expensive to perform such mimicry.

In 2019, Tech Reporter Timothy B. Lee took it upon himself to create a deepfake of his own. For his research, he set a budget of $1,000 to create such a deepfake and chose a popular target, Mark Zuckerberg. Timothy was able to produce a short video deepfaking the face of Star Trek character Data onto Zuckerberg's face, as shown in Fig. 2.3, and it cost him $552 USD ($772 CAD). While the deepfake was of acceptable quality, it was jittery and imperfect, looking unnatural to the point it would be easy to spot even for the untrained eye. Not to mention, there was no vocal cloning for this video, only visual manipulation. Its production was not straightforward either, as Mr. Lee had to manually feed his software, Face Swap, a large amount of training data to procure this result [11].

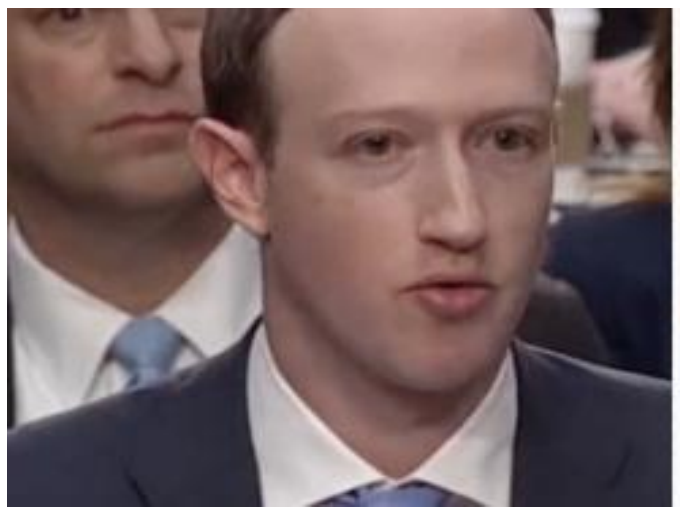

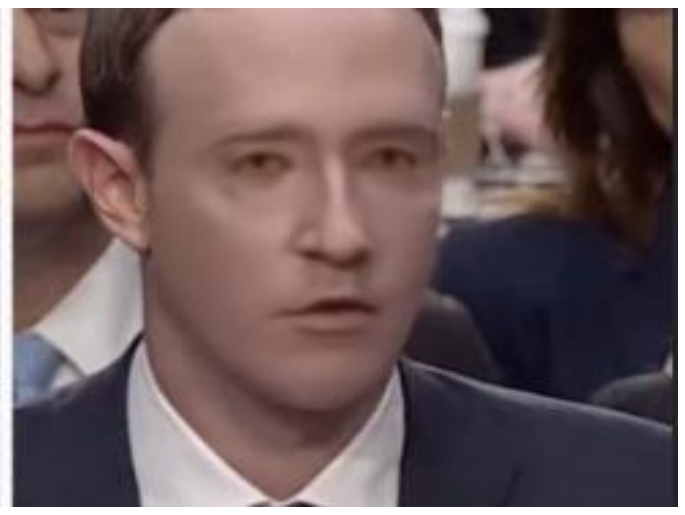

Fig. 2.2. Mark Zuckerberg (top left) and a deepfake of the character Data from Star Trek imposed onto his face (top right). Created by Timothee B. Lee [11].

Only five years later, a superior-looking product, a video over four times as long, with more realistic facial cloning and realistic voice cloning, was produced for less than a third of the deepfake's price, using only easily accessible and easily automated software.

This was not the only deepfake of Mark Zuckerberg created in 2019. In fact, a popular deepfake of Mr. Zuckerberg circulated on the internet, featuring a falsified video and script of him claiming to "control the future with stolen data" [10].

In Fig. 2.4, we can see a side-by-side comparison of this popular deepfake with its victim, as well as a comparison with our deepfake imitation created just over five years later. As you can see, the deepfake below has much cleaner video quality, smoother features, and more accurate facial balance than the deepfake created in 2019.

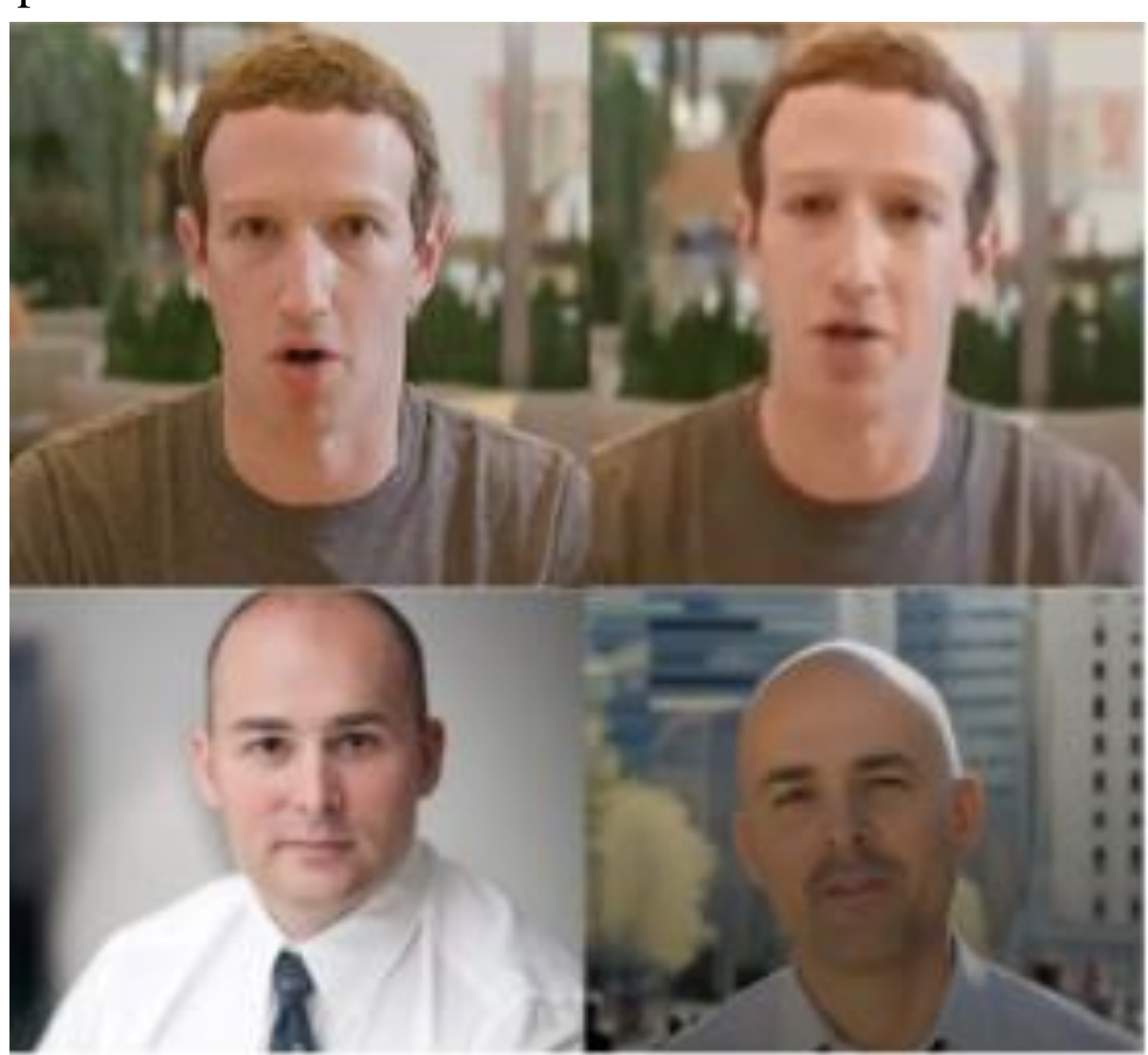

Fig. 2.3. Above: Mark Zuckerberg (top left) and a deepfake replication of his face (top right) (2019) [10]. Below: Claudiu Popa (bottom left) and a deepfake replication of his face (bottom right) (2024).

This level of quality was readily accessible to anyone with an internet connection and could be mimicked for less than $160 CAD. This is especially concerning when looking at the choice of victim. In 2019, it was easy to access hours and hours of both audio and visual references for Mark Zuckerberg, as he is a famously controversial public figure. While our simulated victim, Claudiu Popa, is a CEO and public figure, there is

exponentially less available information about him online. Yet, through our utilization of the openly accessible resources at our disposal, we were able to mimic his appearance and clone his voice nonetheless, with significantly less data.

## III. Deepfake Scenario

### A. Audio-Only

A scammer uses voice cloning technology to impersonate legitimate charity staff during phone calls. They train a voice model to mimic a real staff member's voice and tone accurately by gathering audio samples through publicly available YouTube videos. This cloned voice is used in real-time phone calls to answer questions and build trust with potential victims. Victims are instructed to make donations to fraudulent accounts, believing they are supporting a legitimate cause.

This approach requires fewer resources while maintaining believability. Real-time voice interaction allows for personalized communication, making it more difficult for victims to detect deception. The lack of visual components further enhances the believability of the deception.

Achieving realistic voice cloning with consistent tone, vocabulary, and inflection remains challenging, as latency in voice generation, unusual pauses, or robotic speech can raise suspicion. The scammer must also gather sufficient audio samples of the targeted staff member to train an effective model. Victims familiar with the real staff member may detect inconsistencies in speech patterns, risking exposure of the fraud.

### B. Real-Time Video & Audio

A scammer uses deepfake technology to create a real-time live video feed on Zoom or any other video chat software, impersonating charity staff members. Using AI-powered facial synthesis and voice cloning tools, the scammer appears visually and audibly identical to the legitimate staff. This real-time interaction aims to provide a layer of authenticity for skeptical victims who demand face-to-face confirmation, which is especially apparent in significant financial transactions.

This approach is highly convincing, as it combines visual and audio elements to mimic a genuine interaction. It builds trust among victims by addressing questions in real time and providing a personal touch. The visual component, especially when integrated with realistic lip-syncing and facial expressions, significantly enhances believability.

Real-time video requires advanced AI models and high-performance computing to ensure synchronization of audio and lip movements without noticeable latency. Any mismatched expressions or delays could raise suspicion. Victims familiar with the real staff member may detect inconsistencies in facial features and movements, risking exposure of the fraud.

### C. Pre-Recorded Video & Audio

The scammer uses deepfake technology to create a recorded video featuring an AI-generated celebrity or cloned charity staff member expressing concern for the challenges faced by homeless people. This video is posted on a cloned version of a legitimate charity's website, and then, by using social media, it can be spread to deceive numerous unsuspecting victims.

This approach uses emotional appeal and the social reach of the internet. The recorded video gains attention from viewers, deceiving them with visuals and convincing narratives. The ability to reach a wide audience online enhances the scammer's capacity to persuade victims to donate, using the exposure and engagement generated through internet views and attention to snowball their deception further.

Synchronizing high-quality visuals and audio requires advanced tools and expertise. An active social media account or an account with good standing is required to build trust on the internet. Creating an identical clone of the legitimate charity website and branding, as well as being able to respond effectively to donor queries, adds operational challenges. Cloning the website and managing real-time elements increases the risk of detection once victim suspicion is raised on the internet.

## IV. Deepfake Audio

Deepfake audio technology has become a transformative tool in the field of synthetic speech generation, enabling the creation of highly realistic voice outputs with minimal technical expertise [12]. As this technology becomes increasingly commoditized, it presents a dual-edged potential, offering innovative applications in accessibility, entertainment, and communication while simultaneously posing significant risks, such as facilitating scams, impersonation, and misinformation [13]. This section evaluates the capabilities and implications of deepfake audio tools, focusing on three prominent platforms: Speechify, Play.ht, and ElevenLabs.

### A. Methodology

*1) Selection:* The tools were chosen primarily for their accessibility, as the goal was to evaluate options that are readily available to the public. Accessibility was defined by the ease with which users could sign up, access the tool's features, and use its services without requiring significant technical expertise. While accessibility was the primary factor, pricing was also considered to ensure that the tools selected represented a range of cost options, from free tiers to subscription-based models.

*2) Testing & Evaluation:* To evaluate the tools' voice cloning capabilities, the following methodology was used.

*a) Samples:* Audio samples were sourced from interview recordings posted on the client's YouTube page. These interviews featured the client speaking in varied tones and contexts, providing a diverse dataset for evaluation. The recordings were trimmed to include only the client's parts, resulting in a final 10-minute sample that was used across all three tools to generate comparable outputs.

*b) Criteria:*

- *Realism:* The degree to which the synthesized audio sounded natural and indistinguishable from the sample.
- *Ease of Use:* The simplicity and intuitiveness of the tool's user interface, including the effort required to upload samples, adjust settings, and generate audio.
- *Customization:* The extent of control provided over factors like tone and inflection.

### B. *Comparison*

#### 1) *Obeservations*

*a) Realism:* ElevenLabs produced the most natural and convincing audio, closely replicating the tone, inflection, and nuances of the client's voice. Play.ht followed, producing acceptable results but with slightly less natural tonal variations. Speechify, while functional, produced audio that sounded more robotic and lacked the authenticity of the other tools.

*b) Ease of Use:* Speechify provided the simplest and most straightforward user experience, with an intuitive interface that required no additional configuration. ElevenLabs followed, maintaining an accessible design while incorporating slightly more advanced features. Play.ht, while functional, felt less streamlined and required more effort to navigate compared to the other tools.

*c) Customization:* ElevenLabs and Play.ht both provided robust customization options, with ElevenLabs offering controls for Model Selection (allows choosing from five different voice models), Stability (ensures consistent output but can make speech sound monotone), Similarity (boosts voice clarity and resemblance but may introduce artifacts at high levels), Style Exaggeration (amplifies the expressiveness of speech but may reduce stability), and Speaker Boost (enhances the similarity between synthesized and original voices but may slow down generation speed), as shown in Fig. 4.1.

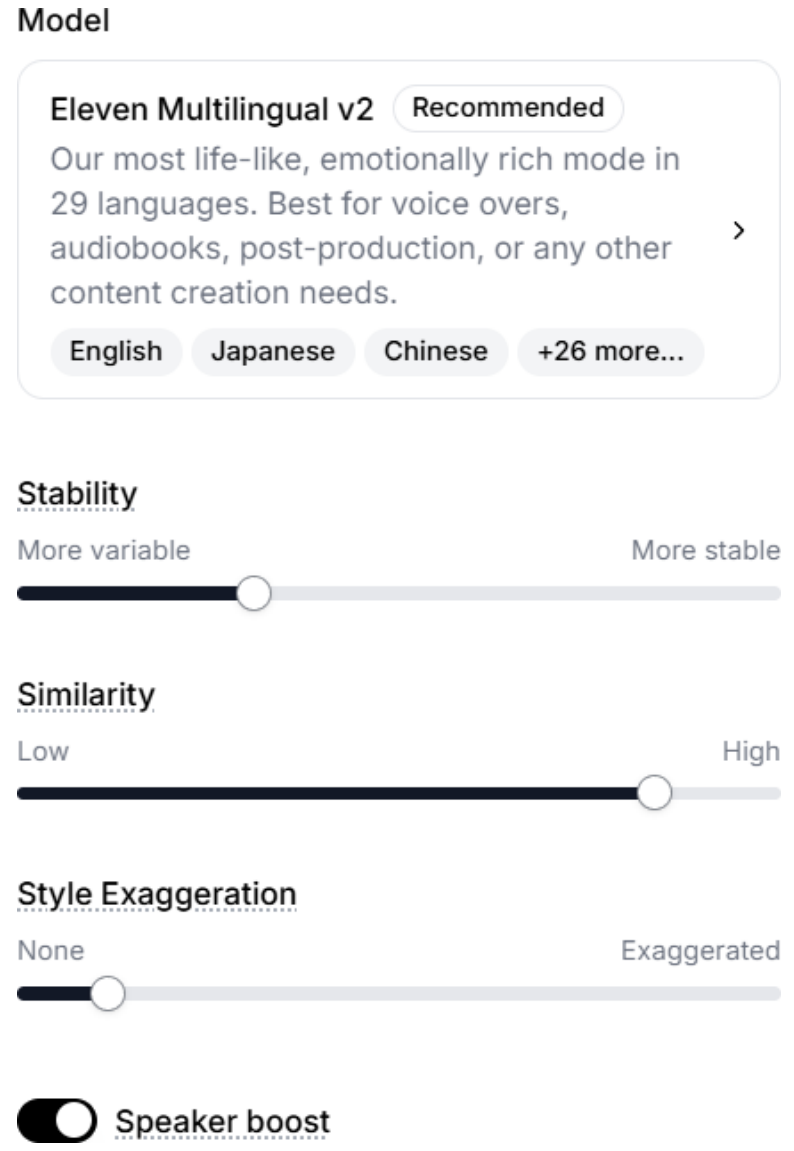

Fig. 4.1. ElevenLabs' configuration options.

Play.ht followed closely, featuring similar controls for Stability, Similarity, and Intensity, as shown in Fig. 4.2.

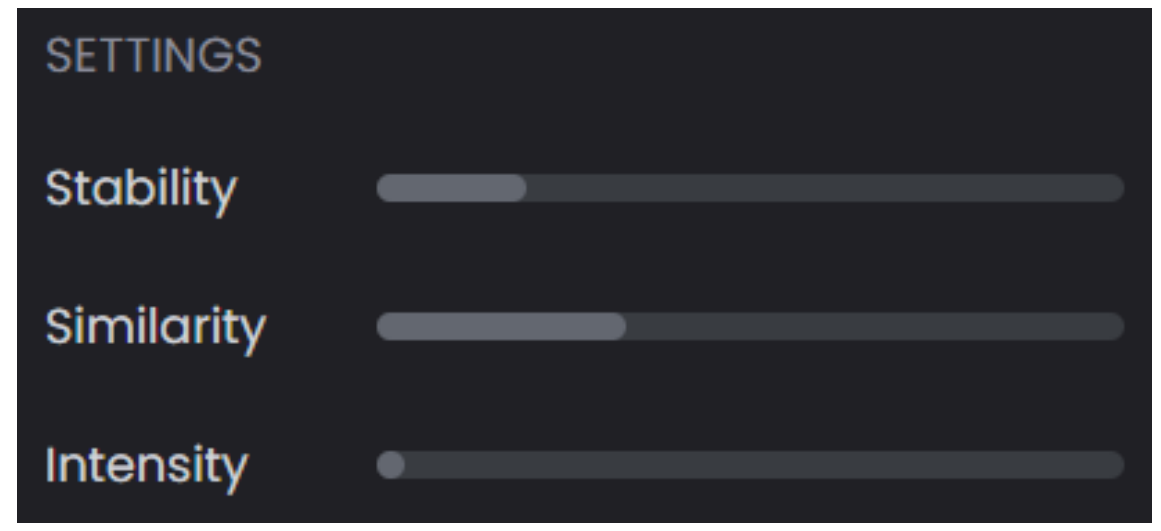

Fig. 4.2. Play.ht's configuration options.

Speechify, however, offered no customization features, making it less adaptable for creating tailored audio.

*2) Verdict:* After thorough evaluation, ElevenLabs was selected for further analysis due to its superior realism and advanced customization options, which offered the highest level of fidelity and control over the synthesized audio. Although Speechify was the easiest to use, its lack of customization features and lower audio quality made it less suitable for this project. Similarly, while Play.ht provided robust customization options, its audio realism and overall user experience fell short compared to ElevenLabs.

### C. *ElevenLabs Analysis*

#### 1) *Features*

*a) Instant Voice Cloning:* Users can clone a voice with minimal input, requiring only a short sample for high-quality results.

*b) Background Noise Removal:* Provides an option to remove background noise from audio samples, eliminating the need for external preprocessing.

*c) Advanced Customization:* Provides controls for Model Selection, Stability, Similarity, Style Exaggeration, and Speaker Boost, allowing precise adjustments to tone, inflection, and fidelity.

*d) Voice Clone Library:* Users can create and manage up to 30 unique voice clones.

*e) Text to Speech:* Users can input prompts to be read in their cloned voice models, replicating the cloned voice with high accuracy.

*f) Speech to Speech/Voice Changer:* Users can record their voice and overlay it with the cloned voice, allowing smooth and natural-sounding audio transformation.

*g) Generation History:* Keeps a record of generated voice lines, allowing users to download past outputs, restore the text for reuse, or recover the settings used for the generation.

*h) Affordable Access:* A starter subscription, priced at $5 USD per month, provides access to instant voice cloning with 30,000 credits (~30 minutes of audio).

#### 2) *Strengths*

*a) Minimal Input Requirements:* Requires only a short audio sample to produce realistic and lifelike audio.

*b) Consistent Performance:* Delivers reliable results across various voice cloning scenarios.

*c) User-Friendly Interface:* Simplifies the process of creating and managing voice models with intuitive tools and real-time previews.

*d) Affordability:* Entry-level subscription starting at $5 USD per month provides sufficient credits for casual users or small-scale projects.

*3) Limitations*

*a) Learning Curve:* Feaures such as Stability and Style Exaggeration may require trial and error to achieve optimal results.

*b) Audio Length:* Starter subscription credits (~30 minutes of audio) may not be sufficient for large-scale projects.

*c) Subscription-Based Pricing:* Although affordable, recurring costs may deter users with limited budgets.

*d) Ethical Concerns:* High-quality cloning raises risks of misuse, such as impersonation, scams, or disinformation.

*4) Applications*

*a) Accessibility:* Assists individuals with speech impairments by generating personalized synthetic voices.

*b) Localization:* Supports multilingual audio generation for adapting content to global audiences.

*c) Education & Training:* Enables realistic simulations, training modules, and voice-based learning tools.

*d) Content Creation:* Suitable for creating voiceovers for podcasts, audiobooks, and videos.

*e) Voice Backup:* Useful for professionals (e.g., voice actors) who want to create and store high-quality clones of their voices for future projects.

### *D. Ethical & Security Implications*

#### *1) Ethical Implications*

*a) Accessbility & Inclusion:* Voice cloning has the potential to make technology more inclusive, particularly for individuals with disabilities. For example, it can help individuals who are at risk of losing their ability to speak by replicating their voice for communication [14].

*b) Authenticity & Trust:* The ability to replicate voices indistinguishably raises concerns about the loss of trust in audio recordings [15]. Deepfake audio could undermine the credibility of authentic content, especially in legal, journalistic, or historical contexts [16].

*c) Consent & Ownership:* Voice cloning without explicit consent violates ethical boundaries, raising questions about voice ownership and intellectual property. In particular, public figures are at risk of having their voices cloned without permission for unauthorized uses [17].

#### *2) Security Implications*

*a) Privacy Violations:* Cloning voices without consent breaches privacy and could lead to identity theft or emotional harm to the individuals affected [18].

*b) Fraud & Scams:* Deepfake audio can be weaponized for fraudulent activities, such as impersonating individuals in phone-based scams, phishing schemes, or financial fraud. For example, cloned voices could be used to impersonate company executives and authorize unauthorized transactions [19].

*c) Disinformation Campaigns:* Deepfake audio could be used to create recordings of public figures, spreading false information or propaganda. This poses a significant threat to political stability, public discourse, and societal trust [20].

*d) Cyberattack Automation:* Deepfake audio technology could be integrated into automated systems for large-scale attacks, such as generating thousands of convincing scam calls simultaneously, increasing the efficiency and impact of malicious campaigns [21].

#### *3) Mitigation Stratigies*

*a) Regulation & Guidelines:* Governments and industry bodies should establish clear regulations on the ethical use of deepfake audio, including requirements for user consent and restrictions on misuse [22].

*b) Technological Safeguards:* Implementing watermarking or traceability features in synthetic audio, like steganography, could help distinguish deepfake audio from authentic recordings [23]. Advanced detection algorithms should be developed and made publicly available to identify deepfake audio in real-time.

*c) Public Awareness & Education:* Raising awareness about the capabilities and risks of deepfake audio is crucial. Educational campaigns can help individuals and organizations recognize potential threats and protect themselves from scams and disinformation [24].

*d) Collaboration Across Sectors:* Collaboration between academic institutions, researchers, industry, and government can lead to shared strategies and resources for addressing the challenges posed by deepfake audio technology [25].

The analysis of deepfake audio tools reveals the fine line between their potential for innovation and the risks of misuse. ElevenLabs, with its superior realism and customization features, demonstrates how far voice cloning technology has advanced, making it a valuable tool for applications ranging from accessibility to education. However, the ease of creating highly realistic audio also increases concerns about fraud, privacy breaches, and ethical misuse. As deepfake audio continues to evolve, it is crucial to implement safeguards such as regulations, detection mechanisms, and public awareness to mitigate its potential for harm.

## V. Deepfake Video

"Deepfake" (or its plural "deepfakes") is defined by Merriam-Webster Dictionary as "an image or recording that has been convincingly altered and manipulated to misrepresent someone as doing or saying something that was not actually done or said" [26]. In recent years, deepfakes have gained significant traction due to rapid advancements in AI technologies. This surge in popularity is also fueled by the influx of AI generative tools such as ChatGPT, Gemini, Runway, and numerous others that are readily available online, ranging from free to paid, making it accessible en masse. Our aim was to find a way to create realistic deepfakes using openly available tools that are not resource-intensive to showcase the fact that realistic deepfakes can nowadays be created by almost anyone with a few hundred dollars to their name. During our research, we tested multiple tools such as DeepLiveCam [27], SimSwap [28], and Rope [29]. We ultimately decided to utilize Rope NEXT [30], which is a fork of the original Rope software created by Hilobar due to its performance, ease of use, and customizability.

### A. *Rope Analysis*

Rope is a fork of an open-source face-swapping software, originally named Roop, created by Somdev Sangwan [31]. The software runs on Python and enables users to swap faces on pictures and videos without the need for any training and/or using a dataset. The only requirement for doing a face swap using Rope is to have only one image of the desired face. This makes Rope unique, as it eliminates the need to train on a dataset, providing a user-friendly GUI with different customizability, all the while providing high-quality results as seen in Fig. 5.1.

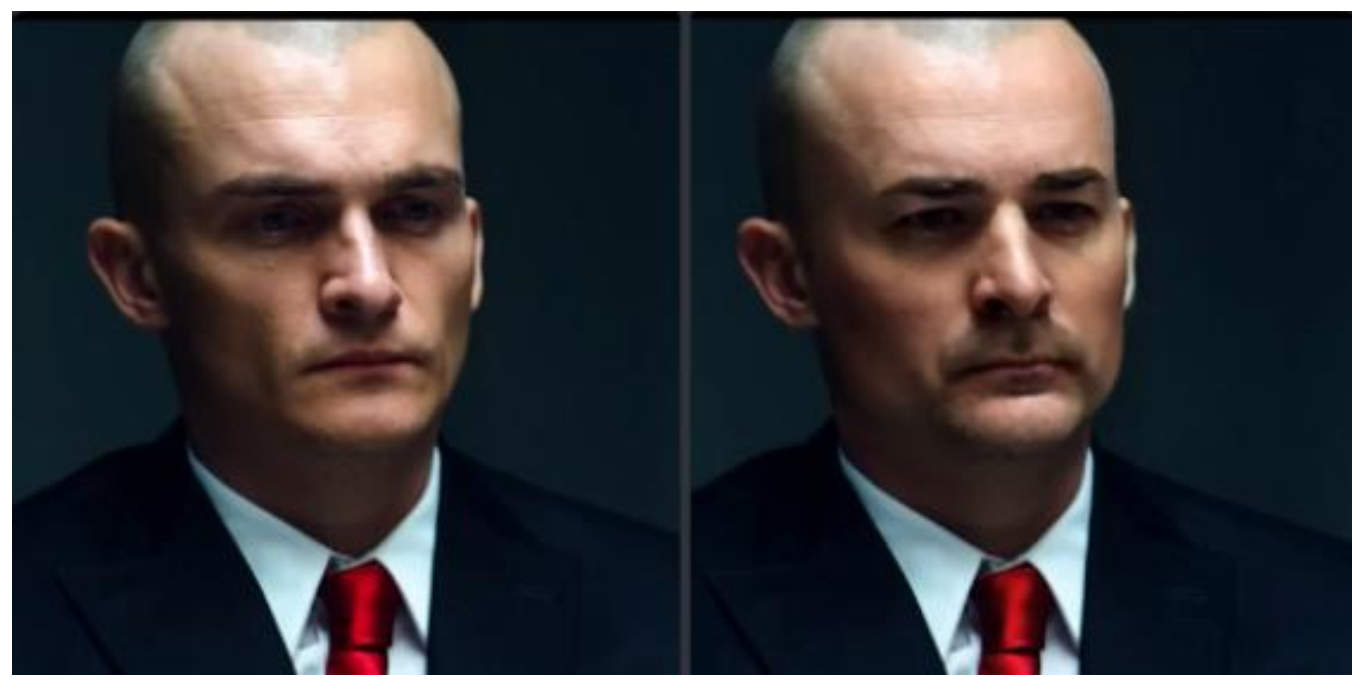

Fig. 5.1. On the left, we see Rupert Friend as Agent 47 from the movie Hitman (2015) [32] and on the right is Claudiu Popa, our project sponsor, face swapped into the clip using Rope.

### B. *Testing & Evaluvation*

We configured Rope on a system with the following specifications:

- CPU: AMD Ryzen 7 5800H 3.2 GHz
- GPU: Nvidia GeForce RTX 3060 laptop 6 GB
- RAM: 8+8 GB 3200mhz
- DISK: WD Black SN770 1TB SSD

There are quite a few settings and parameters that can be altered to create an optimal deepfake, the most important one in our usage is the "restorer" option, which increases the resolution of the swapped face. Originally, with the default parameters, the faces once swapped look grainy and not clear at all. What face restorer does is enhance the final swapped face to make it look clear and not grainy or pixelated.

The restorer has several options to select from. Here is the list of them from the least to the most time consuming:

- GPEN256
- GFPAN
- CF
- GPEN512

To test the differences between the different restorer models, we decided to use a 4-second video, apply face swaps using the different restorer models, and record the render time as seen in Table II, along with the stills of the swapped video for each restorer in Fig. 5.2.

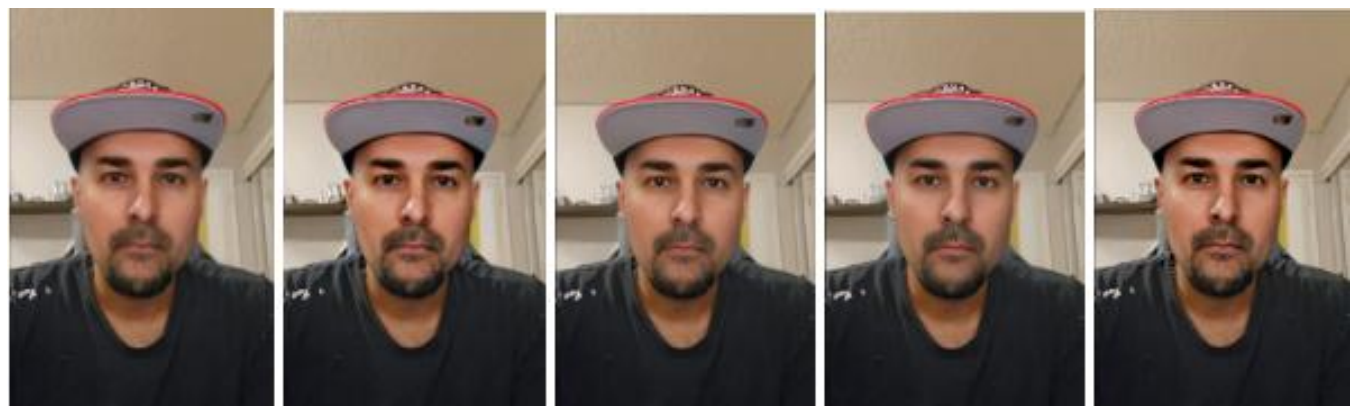

Fig. 5.2. (From left to right) No restorer, GPEN256, GFPGAN, CF, GPEN512.

TABLE II. RESTORER MODEL COMPARISONS (ROPE)

| **Restorer Model Name** | **Render Time (seconds)** |
| --- | --- |
| No Restorer | 18 |
| GPEN256 | 23.8 |
| GFPGAN | 38.2 |
| CF | 53.3 |
| GPEN512 | 60.7 |

### C. *Challenges*

*1) Hair, Beard & Skin Color:* We noticed that Rope only seems to apply the face and not the hair or beard of the input/desired face. This makes it harder for us to create deepfakes of our sponsor, who is bald, as it transplants the hair from the original video to the input face. This problem could also be reversed if the input/desired face had a unique beard/hair style that'd be hard to match. To combat this, we have thought of three easy strategies that we can implement.

*a) Head Coverings:* The easiest one to implement by far is to record original videos with hoodies or hats on, as seen in Fig. 3.2, which eliminates the need to worry about hairstyles but still poses an issue with unique beard styles, as it seems to blend the two facial hair styles together.

*b) Accessories:* The other option would be to wear bald caps, wigs, and fake facial hair. While this option is certainly possible, we personally have not tried it for this project.

*c) AI Generated Content:* The last strategy would be to use AI-generated people with the same physical description as our sponsor. Using text-to-video or text-to-image generators, we can effectively create videos according to our needs, as well as do things we would not be able to do otherwise, like a deepfake video of our sponsor doing a backflip, for example (something that is tremendously difficult physically). This also solves our primary concern, which is the different hairstyle, beard style, and skin and hair color.

*2) Computing Resources:* The program occasionally encountered performance challenges, specifically issues such as freezing and stuttering, when attempting to swap using more complex restorer models and/or adjusting other settings. This limitation was directly tied to our 6 GB video RAM capacity, which also contributed to longer rendering times. To combat this issue, we decided to purchase an Amazon EC2 instance with more graphical computational power than we had previously at our disposal. We did give this a try, but unfortunately, due to other issues such as the installation not being very user-friendly and suitable for Linux systems, we decided to abandon the utilization of a virtual machine and explore other options. As we were researching options to optimize Rope, we found a fork of Hilobar's Rope made by Alucard24 called Rope NEXT. The main difference between Hilobar's Rope and Rope NEXT is the implementation of Nvidia's TensorRT, which sped up the process significantly, allowing more computationally intensive restorer models to run quickly even with the minimal 6GB video RAM on the RTX 3060. To show you the performance gain, Table III shows the render time for the same video demonstrated in the earlier section using Rope NEXT.

TABLE III. RESTORER MODEL COMPARISONS (ROPE NEXT)

| **Restorer Model Name** | **Render Time (seconds)** |
|---|---|
| No Restorer | 7.1 |
| GPEN256 | 7.5 |
| GFPGAN | 12.8 |
| CF | N/A |
| GPEN512 | 13.3 |

When comparing Table II and Table III, we can see that the performance of Rope NEXT is much higher than Rope, with performance improvements of up to 4.6 times that of Rope in certain cases.

### D. Cause for Concern

Photoshop and CGI have existed for quite a long time, and these tools have been used over the years to manipulate and create realistic fakes, but not "deepfakes," as deepfakes are created with the aid of artificial intelligence [33] and therefore are undecipherable to human eyes and even to an AI model due to the nature of Generative Adversarial Networks, which are widely used in deep learning to perfect deepfakes [34].

As demonstrated, open-source software such as Rope, once set up, is child's play to use in creating deepfakes. Photoshop and CGI require a lot of knowledge and skill in different tools, software, and methods, while creating deepfakes using Rope only requires the user to specify the video they want to apply the deepfake on and what face to use. This ease of use may lead to an influx of deepfake videos on the internet, especially due to the numerous large social media and video-sharing platforms that can be used for propagation, free of charge [35].

## VI. GENERATIVE AI

The commoditization of generative AI and its widespread availability, combined with its affordability, has drastically changed the technological landscape today, and in the future, it will drive creativity and innovation across industries. However, its misuse and rapid adoption will also undermine trust and create ethical, legal, and societal challenges that must be addressed. Tools like Runway, DALL-E, and LTX Studio empower users to create high-quality synthetic media with minimal input, lowering the barriers to entry to access this technology. The duality of generative AI is particularly evident in its application to deepfake technology, where it can serve both constructive and destructive purposes.

Generative AI introduces an alarmingly extensive range of ethical challenges which includes privacy, misinformation, and ultimately exploiting societal inequality. The groundbreaking potential that this technology showcases requires rigorous evaluation of its effects to ensure responsible use in the digital world [36].

### A. Applications in Deepfake Technology: Opportunities & Risks

Deepfake technology emphasizes the dual nature of generative AI. Originally developed for entertainment and creative expression, deepfakes have evolved into versatile tools that can educate, deceive, or disrupt.

Generative AI deepfakes offer numerous positive applications:

*1) Cybersecurity Training*: Simulating thousands of different scenarios and tailoring each scenario to specific recipients based on who would be most susceptible to each unique scenario can showcase phishing attacks and impersonation scenarios that help individuals recognize and respond to these threats.

*2) Awareness Campaigns*: Demonstrating the dangers of deepfakes educates the public on the technology's implications, which raises a question for the future: will individuals be able to trust that the media they consume is authentic and not artificially generated?

*3) Creative Storytelling*: Filmmakers can utilize deepfakes to create realistic stories, enhancing storytelling capabilities by incorporating historical figures or well-known personalities from the past into modern or futuristic sources of media for the public to enjoy.

### B. Risks: Exploitation for Harm

The same technology can be exploited for malicious purposes:

*1) Disinformation*: Generating convincing fake news or political speeches undermines public trust and manipulates opinion [43].

*2) Erosion of Trust in Digital Evidence*: The reliability of future digital evidence in courts and investigations is increasingly called into question. This raises a critical issue: can we trust digital evidence when artificial content could now and in the future be generated to look identical to authentic media? This undermines not only the legal system but also industries like law enforcement and insurance, where authenticity is critical [45].

*3) Non-Consensual Content*: Creating explicit materials featuring individuals without their consent raises significant ethical and legal concerns, yet it remains the biggest driving factor of the growth in deepfake technology.

### *C. Challenges with Hair Movement in Deepfakes*

A technical challenge we encountered in our deepfake creation is the accurate rendering of moving elements, notably hair movement. Hair, with its constantly shifting individual movement, poses difficulties in image synthesis. Traditional deepfake tools often fail to synchronize this hair motion with facial expressions across multiple frames, which leads to noticeable imperfections that take away from the realism of the media.

*1) Our Experience:* During our project, hair movement emerged as a persistent issue, as seen below in Fig. 6.1. Despite experimenting with several tools, we encountered unnatural hair dynamics that distanced themselves from the overall authenticity of the deepfake. This challenge required the adoption of more advanced generative AI solutions capable of synchronizing hair and other moving elements seamlessly. Studies like those by Unite AI [39] highlight how hair behavior remains one of the most troublesome aspects of synthetic media creation.

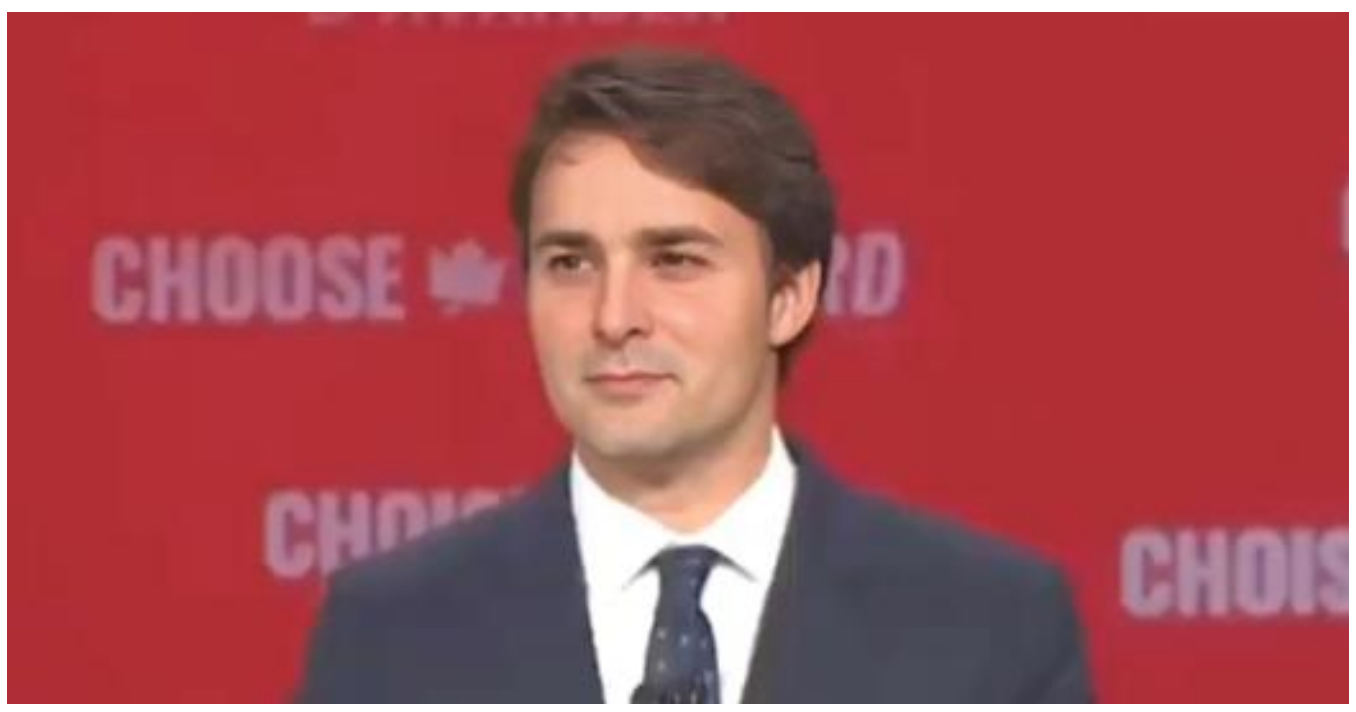

Fig. 6.1. Deepfake test with inaccurate hair.

### *D. Adoption of Generative AI for Enhanced Realism*

Generative AI models that are trained on large-scale datasets of multiple different images and videos have proven effective in addressing the challenges of hair movement. These models can generate content that maintains the same quality across an image or video, ensuring that all moving elements, including hair, are consistently integrated together to increase realism. The result is a more realistic and convincing deepfake that surpasses the limitations of traditional techniques [39].

Runway's Gen-3 Alpha offers significant improvements in this area of motion consistency and realistic detail. Unlike other tools such as LTX Studio or Canva, which either lack realistic details or rely heavily on stock content, Runway's Gen-3 Alpha introduces 3D dynamics into generative video production [42]. Although Runway does come at a significant cost, making it less accessible to people creating smaller projects compared to free or low-cost alternatives, it offers higher-quality results.

The flowchart, as seen below in Fig. 6.2, outlines the step-by-step process used in our project to generate deepfake videos. This workflow illustrates how various generative AI tools were iteratively applied to ensure optimal quality, including text-to-video generation, video extension, and synchronization using ROPE and lip-sync technologies built into Runway.

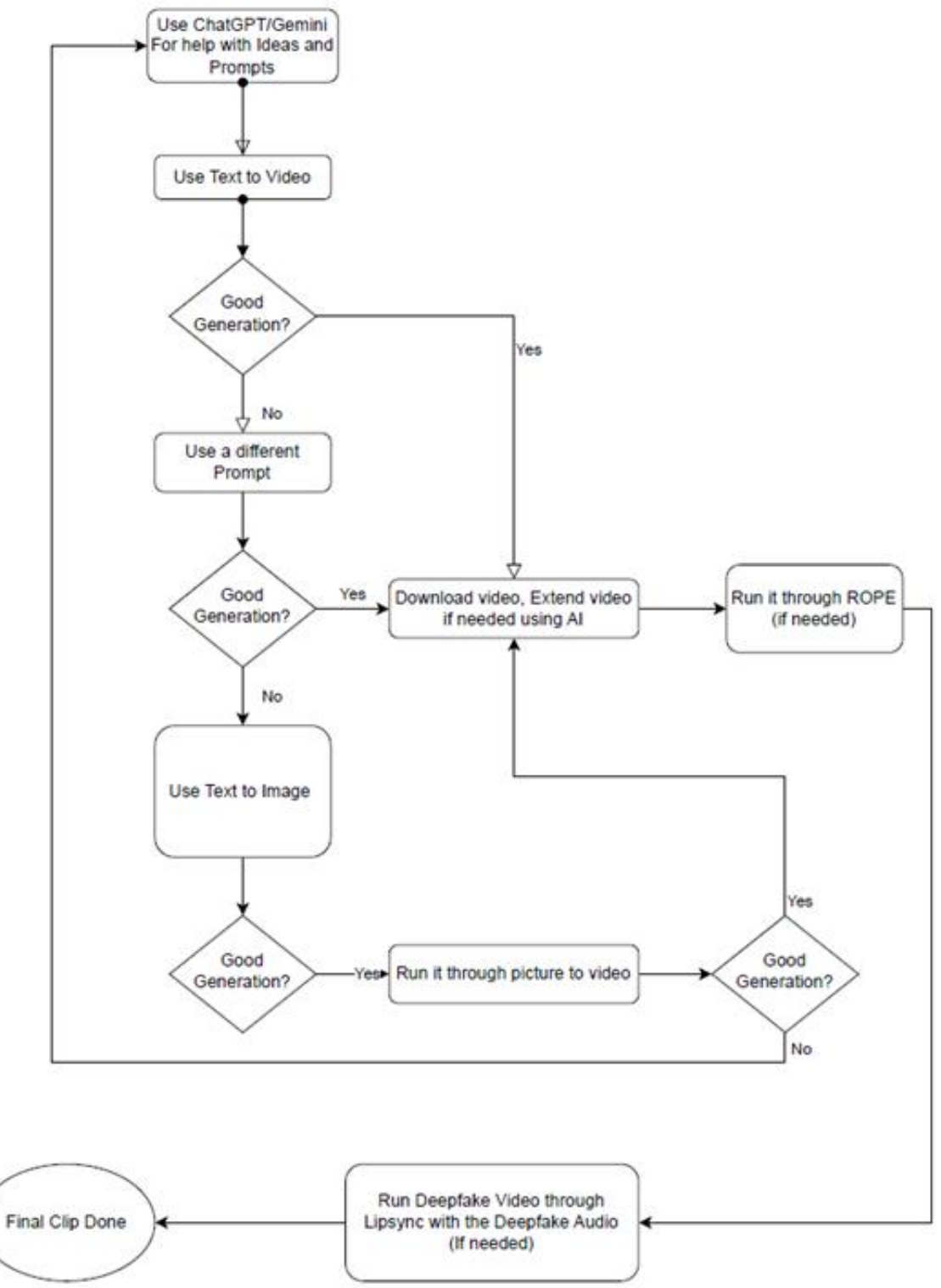

Fig. 6.2. Deepfake flowchart.

### *E. Advancements in Generative AI Models*

Generative AI models have significantly advanced content creation by enabling machines to produce data that closely resembles real-world inputs. Among these models are Generative Adversarial Networks (GANs) and Variational Autoencoders (VAEs), both of which have been instrumental in generating realistic images, videos, and audio.

*1) Generative Adversarial Networks (GANS):* GANs consist of two neural networks working together: one makes

fake data (generator), and the other decides if the data is real or fake (discriminator). They compete, and over time, the generator gets better at creating data that looks real. This process enables the creation of highly realistic synthetic data. GANs have been widely used in applications ranging from image synthesis to video generation, showcasing their versatility in handling complex tasks [42]. Although GANs pose significant risks to the legal system, the technology's capacity to generate convincing fake evidence, such as manipulated videos or documents, creates countless scenarios where legal cases might hinge on false information. As GANs continue to improve in the future, their misuse can undermine trust in digital evidence and judicial processes.

*2) Variational Autoencoders (VAE):* VAEs are generative models that compress input data into a simplified representation called a latent space, capturing the most critical features. This latent space represents the data as a distribution, and VAEs sample random points from this space to create new, realistic variations of the original data [42]. This process allows VAEs to generate diverse and smooth outputs from existing data, making them especially useful for applications like generating images or audio with smooth transitions. VAEs' ability to produce plausible variations of data can be misused for subtle manipulations both in voice cloning and altering audio recordings to falsify conversations.

### *F. Runway's Gen-3 Alpha Model*

Runway's Gen-3 Alpha is a huge leap forward in the generative AI space, notably in video synthesis. This model uses a combination of visual transformers and diffusion models to achieve generated videos that can be hard to differentiate from reality [40].

*a) Visual Transformers:* These models look at the relationships within data, from pixels in an image to frames in a video and use this information to generate corresponding outputs. They are effective at looking at extended amounts of patterns, which is important for maintaining a consistent feel and narrative across multiple frames in video synthesis.

*b) Diffusion Models*: These models start from completely distorted data and then reconstruct that data step by step, refining it each time to get closer to a refined set of data until it results in highly detailed and accurate outputs.

One of the most significant developments in generative AI tools, such as Runway's Gen-3 Alpha, is its capacity to generate real-life visuals from a single prompt and expand on that image to produce an authentic video scenario. While these capabilities make it a valuable tool for creative endeavors, deepfake generation, and AI-enhanced storytelling, they also significantly increase the risks of misuse. The ability to create persuasive and captivating media raises concerns about its use in creating fraudulent content, such as forged evidence or manipulated videos. These risks further emphasize the need for stricter regulations and ethical guidelines to regulate the use of advanced generative AI technologies.

Fig. 6.3 below exemplifies this capability, showcasing how Runway can generate synthetic portraits with intricate details, realistic lighting, and dynamic backgrounds. In this example, the generated image showcases subtle features like facial expressions and environmental depth. Such outputs highlight Runway's suitability for creating high-quality synthetic media and the eventual dangers of this technology.

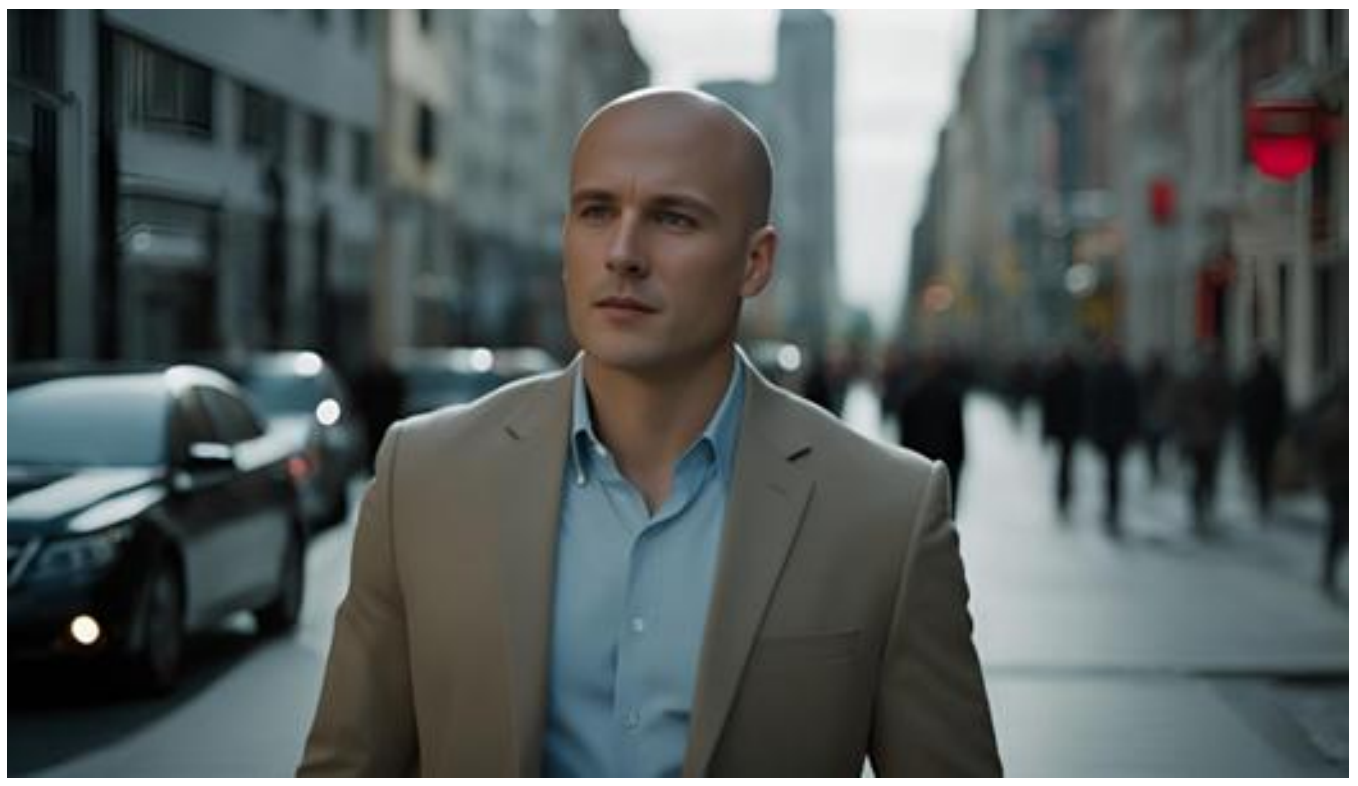

Fig. 6.3. AI-generated portrait using Runway's Gen 3 Alpha.

### *G. Evaluation of Generative AI Tools*

TABLE IV. COMPARISON TABLE

| Tool | Strengths | Weaknesses |
|---|---|---|
| LTX Studio | Generates basic scenarios. | Limited accuracy, poor mouth movements. |
| vivago.ai | Generates synthetic videos. | Long processing time, low quality, facial deformities. |
| invideo.io | Stock-based video creation. | Reliance on stock content, irrelevant narrations. |
| Canva | Accessible, mid-quality videos. | Watermarked, requires premium for full features. |
| Runway | Advanced 3D dynamics, high realism. | High computational cost, subscription fee. |

## VII. CONCERNS AND IMPLICATIONS

After being able to create realistic and believable deepfakes with relative ease and low cost, we believe that video, photo, and audio are in danger of losing their authenticity that we traditionally associate them with. As mentioned earlier, the problem is not just with the manipulation of said content beyond the distinguishable threshold of the human eye, but rather the way deepfakes work. GANs, which are a vital part in the creation of deepfakes, enable the creation of deepfakes that are more accurate and authentic than the original content, making it more challenging or impossible for AI detection tools to viably detect a deepfake as it is made to be indistinguishable from the original content in the first place.

With the high influx of generative AI and deepfake tools that allow for the creation of highly realistic and convincing deepfakes at a relatively low cost and effort, we face an emerging threat to the authenticity of video, photo, and audio content, which has been held as the pinnacle of evidence in differentiating fact and fiction. The problem extends beyond the manipulation of content to a level indistinguishable to the human eye. It lies in the very vital mechanism of deepfakes: Generative Adversarial Networks (GANs), a cornerstone of deepfake technology. GANs can produce content that is often more precise and authentic looking than the original content. This sophistication poses a significant challenge for AI detection tools, as the goal of deepfakes is to be virtually indistinguishable from genuine media, rendering detection increasingly unviable. This problem is so genuine and worrisome that DARPA is actively looking at ideas and ways to detect deepfakes and verify the authenticity of content through their Semantic Forensics program [46], which aims to achieve commercialization of deepfake defense technologies, as well as their AI Forensics Open Research Challenge Evaluation (AI FORCE) program, which aims to create an open-source research platform to further develop defenses against deepfake threats.

While deepfakes can be detected by the utilization of AI models, we can protect the authenticity of content using blockchain technology [47], as mentioned by Gambin et al., by utilizing the concept of distributed, tamper-proof ledgers that record the information regarding the content, which can also be validated and audited, similar to a cryptocurrency transaction. This centralized system can be utilized in a way to preserve the authenticity of content once challenges involving the practicality and implementation of such systems on a wide scale can be sorted [48].

As forks of Rope and other novel software programs similar to Rope keep being worked on and improved upon, while remaining accessible to all, it's only a matter of time until we are unable to differentiate fact from fiction unless sufficient progress is made to keep up with the various new AI models and programs utilized in the creation of deepfakes, either through detection of deepfakes or content authentication.

## VIII. Conclusion

The commoditization of deepfake technology poses both a profound opportunity and a significant risk to society. As demonstrated, its applications range from empowering creators to revolutionize media production, to enabling malicious actors to undermine trust and security. The increasing accessibility of deepfake tools demonstrates the transformative potential of this technology in industries such as entertainment, marketing, and communication. However, it also highlights the need for vigilant regulation, ethical considerations, and security measures to mitigate the risks of its misuse. This white paper highlights the importance of balancing innovation with responsibility, advocating for a framework that allows for the positive use of deepfakes while safeguarding against harm. As deepfake technology continues to evolve, it is crucial for policymakers, technologists, and society at large to work collaboratively to ensure its safe and ethical integration into our digital landscape.